\documentclass[12pt]{article}
\usepackage[dvips]{graphicx}
\usepackage{amssymb}
\usepackage{amsmath}
\usepackage{epsfig}
\usepackage{cite}
\usepackage{hyperref}
\usepackage{bbold}
\usepackage{multirow}

\numberwithin{equation}{section}
\numberwithin{table}{section}

\def\beq{\begin{equation}}
\def\eeq{\end{equation}}
\def\be{\begin{equation}}
\def\ee{\end{equation}}
\def\bea{\begin{eqnarray}}
\def\eea{\end{eqnarray}}

\def\jbar{\bar \jmath}
\def\ibar{\bar \imath}

\DeclareRobustCommand{\SkipTocEntry}[4]{}
\RequirePackage{color}

\newcommand{\cO}{\mathcal{O}}

\newcommand{\dd}{\mathrm{d}}

\begin{document}

\begin{titlepage}
\begin{center}
\rightline{\small }

\begin{flushright} 
CPHT-RR107.122019 \\
IPhT-T19/167
\end{flushright}

\vskip 2cm

{\Large \bf An update on moduli stabilization \\with  antibrane uplift}
\vskip 1.2cm

Emilian Dudas$^{\dag}$ and
Severin L\"ust$^{\dag,*\ }$ 
\vskip 0.1cm
{\small\it  $^\dag$ Centre de Physique Th\'eorique, Ecole Polytechnique, CNRS \\ 
91128 Palaiseau Cedex, France} \\
\vskip 0.1cm
{\small\it  $^{*}$ Institut de Physique Th\'eorique, 
Universit\'e Paris Saclay, CEA, CNRS\\
Orme des Merisiers \\
91191 Gif-sur-Yvette Cedex, France} \\

\vskip 0.8cm

{\tt }

\end{center}

\vskip 1cm

\begin{center}  {\bf Abstract }\\

\end{center}

It was recently shown that in warped compactifications based on a Klebanov-Strassler throat there is a light complex structure field,  governing the size of the throat and the redshift at its tip. 
We show that after uplift of the cosmological constant by an anti-D3 brane at the tip of the throat, the contribution to supersymmetry breaking coming from the new light field is large. 
 We work out the mass scales, in particular the condition for this  field to be heavier than the K\"ahler modulus. We check that for the range of parameters relevant for the destabilization we find
 agreement with de Sitter swampland conjecture.  
Adding matter fields on distant branes, we discuss the effects on supersymmetry breaking in the observable sector.  A hierarchically small scale of supersymmetry 
breaking translates generically into large values of localized D3 charges in the manifold. 


\vspace{0.2cm}

\noindent

\vfill

\noindent
December 2019

\end{titlepage}



\section{Introduction}

The KKLT construction of moduli stabilization \cite{KKLT} relies on a three step procedure. In the first step, fluxes in the internal manifold stabilize all moduli fields, except the overall volume (K\"ahler) modulus $\rho$.\footnote{Here we consider only models with one K\"ahler modulus, i.e.~$h^{1,1}=1$. In more general terms only the complex structure moduli get stabilized by fluxes.}
In a second step, nonperturbative effects like stringy instantons or gaugino condensation on D7 branes stabilize $\rho$ in anti-de Sitter space. Finally, an $\overline{D3}$ 
antibrane is introduced which breaks supersymmetry.\footnote{Actually realizing it nonlinearly, similar to perturbative string constructions of the ``Brane Supersymmetry'' type \cite{bsb}), 
whose nonlinear supersymmetric actions were constructed in \cite{dm}.}  The second and third step were debated over the years \cite{us,tachyons,Joe-Andrea,thomas,sav-ander} and whereas 
there is not yet a consensus over the last step, there are recent positive results indicating the validity of the second step \cite{Hamada:2018qef}.  

Recently \cite{Bena:2018fqc, Blumenhagen:2019qcg}  the first step was addressed in the Klebanov-Strassler (KS) deformed conifold construction, which contains all ingredients needed in the KKLT construction. 
 It was shown that one of the complex structure fields, called $S$ in what follows, which governs the size of the KS throat,  is much lighter than previously thought. Its scalar potential is therefore shallow and it is significantly modified by the uplifting  $\overline{D3}$ antibrane.  Not destabilizing the throat requires a minimal value of one of the flux quanta $M$ \cite{Bena:2018fqc, Blumenhagen:2019qcg, Bena:2019sxm}, 
\begin{equation}
g_s M^2 \ \geq  (6.8)^2  q  \ , \label{intro1}
\end{equation}
where $g_s$ is the string coupling and $q$ is the number of antibranes. In this paper we consider the most favorable case $q=1$.%
\footnote{It can be argued that there could be further corrections to the potential for $S$ which become relevant if one goes far away from the original, supersymmetric minimum.
These corrections could potentially be so important that the bound \eqref{intro1} gets modified significantly or even becomes invalid.
Resolving these issues is beyond the scope of this paper.
Instead, we assume that the results of \cite{Bena:2018fqc, Blumenhagen:2019qcg} are correct.
Moreover, we will mostly work in a regime where \eqref{intro1} is satisfied and it is therefore reasonable to expect that further corrections to the potential are subleading in the region of interest.
}

The purpose of the present letter is to investigate in more detail the consequences for the KKLT construction: the resulting vacuum structure and mass scales, various contributions to supersymmetry 
breaking and the needed localized D3 charge in the internal space which produce physically motivated hierarchies.
 In Section~\ref{sec:KS} we review the effective action for the light complex structure field $S$ in the KS geometry and the mechanism behind the potential destabilization of the throat, once one adds the antibrane uplift. 
 Section~\ref{sec:4D} proposes a 4d supergravity description of the system including the KKLT sector of moduli stabilization and discusses the vacuum structure and supersymmetry breaking. 
 We use a manifestly supersymmetric four-dimensional supergravity description and describe the uplift via
a nilpotent chiral multiplet  in supergravity \cite{nilpotent1,nilpotent2}. 
 In Section~\ref{sec:softterms} we add matter fields and study the effects of supersymmetry breaking, from a 4d perspective and, alternatively, from a higher-dimensional one. 
 We conclude with some comments and a short Appendix. 
      

\section{The effective action of the Klebanov-Strassler warped compactification }
\label{sec:KS}

The traditional KKLT construction of moduli stabilization \cite{KKLT} is based on warped compactifications of Calabi-Yau manifolds, with a constant dilaton, five and three-form fluxes \cite{GP}. 
The background metric and five-form flux are
\begin{equation}\begin{aligned}\label{eq:warpedbackground}
\dd s^2 &= H^{-1/2} \dd s^2_4 + H^{1/2} \dd s^2_{6} \,, \\
F_5 &= \left(1+\ast\right) \mathrm{vol}_4 \wedge \dd H^{-1} \equiv \ast {\cal F}_5+ {\cal F}_5 \,,
\end{aligned}\end{equation}
where $H$ is the warp factor and $\dd s^2_{6}$ is the unwarped metric of the internal manifold. As argued in \cite{GKP}, one can interpret this manifold as a throat-type region of strong warping,
analogous to Randall-Sundrum type models \cite{rs},
 glued to a compact Calabi-Yau space. 
 
 In the region of strong warping the local internal geometry is that of the deformed conifold, defined by its embedding into \(\mathbb{C}^4\),
$\sum_{a = 1}^4 \omega_a^4 = S $. 
The deformation parameter $S$ is the complex structure modulus whose absolute value corresponds to the size of the 3-sphere at the tip of the cone.
The other complex structure moduli $Z^I$ come from the ``UV" geometry. We thus have $h^{2,1}+1$ A-cycles:
\beq\label{eq:S}
\int_{A} \Omega_3= S \ , \qquad  \int_{A_I} \Omega_3= Z^I
\eeq
where $I=0,...,h^{2,1}-1$. We assume that the prepotential splits according to
\begin{equation}
F(S, Z^I) = F_{cf}(S) + F_{UV}(Z^I) \,,
\end{equation}
where $F_{cf}$ is the prepotential of the deformed conifold and  the ``UV prepotential,'' \(F_{UV}\), does not explicitly depend on \(S\).
We thus have
\beq\label{eq:GS}
\int_{B} \Omega_3 = F_S= \frac{S}{2 \pi i}\left( \log{\frac{\Lambda_0^3}{S}} +1 \right)+ F^{0}_S  \, , \qquad \int_{B_I} \Omega_3 = F_I \,,
\eeq
where $F_S$  and $F_I\) are the derivatives of \(F\) with respect to $S$ and $Z_I$ respectively, and $F_S^{0}$ depends on the details of the compactification manifold, but is independent of $S$. The cutoff \(\Lambda_0\) corresponds to the transition between the highly warped region, modeled as a KS throat, and (relatively unwarped) rest of the compact Calabi-Yau manifold.
 
The 3-form fluxes on the 3-cycles are\footnote{The setup only requires one type of flux on each cycle.}
\begin{equation}\begin{aligned}\label{eq:fluxes2}
\frac{1}{(2\pi)^2 \alpha'} F_3 &= M \alpha + M^I \alpha_I - M_I \beta^I \,, \\
\frac{1}{(2\pi)^2 \alpha'} H_3 &= - K \beta  + K^I \alpha_I - K_I \beta^I  \,.
\end{aligned}\end{equation}
where $\alpha_I, \beta^I$ are Poincare duals to the cycles $B_I, A^I$ and we have singled out the RR flux on the $S^3$ cycle at the tip of the throat, $M$, and its NSNS partner $K$. These are the fluxes responsible for the deformation of the conifold by the parameter $S$.

The throat region is that of the Klebanov-Strassler (KS) solution\cite{KS}, with the six-dimensional metric of the deformed conifold\footnote{Note that taking ${\cal T}$ and \(g^i\) to be dimensionless requires the deformation parameter \(S\) to be of dimension \((\mathit{length})^3\).}
\begin{equation}\begin{aligned}\label{eq:defconifoldmetric}
\dd s^2_6 = \frac{\left|S\right|^{2/3}}{2} \mathcal{K}({\cal T}) \biggl[\frac{1}{3 \mathcal{K}^3({\cal T})} \left(\dd {\cal T}^2 + (g^5)^2\right) + &\sinh^2({\cal T}/2)\left((g^1)^2 + (g^2)^2\right) \\ &+ \cosh^2({\cal T}/2)\left((g^3)^2 + (g^4)^2\right)\biggr] \,,
\end{aligned}\end{equation}
where \(g^i\) is an orthogonal basis of one-forms on the base of the cone and
\begin{equation}
\mathcal{K}({\cal T}) = \frac{\left(\sinh(2{\cal T}) - 2{\cal T}\right)^{1/3}}{2^{1/3}\sinh{\cal T}} \,.
\end{equation}
The warp factor of the KS solution is
\begin{equation}\label{eq:warpfactor}
H({\cal T})= 2^{2/3} \frac{{g_s (\alpha' M)^2  (\rho + \bar \rho)}}{\left|S\right|^{4/3}} I({\cal T})
\end{equation}
where
\begin{equation}\label{eq:I}
I({\cal T}) = \int_{\cal T}^\infty \dd x \frac{x \coth x - 1}{\sinh^2 x} \left(\sinh(2x) -2x \right)^{1/3}.
\end{equation}
The UV cutoff $\Lambda_0$ where the solution is glued to the compact Calabi-Yau solution is such that the total NSNS flux over the $B$ cycle is $K$, according to \eqref{eq:fluxes2}:
\beq\label{eq:Kcutoff}
K= \frac{1}{(2\pi)^2 \alpha' } \int_{B} H_3=  \frac{1}{(2\pi)^2 \alpha' } \int_{{\cal T} \le {\cal T}_{0}} \int_{S^2} H_3  \ , \quad \ \Lambda_0^2=\frac{3}{2^{5/3}} \left|S\right|^{2/3} e^{2 {\cal T}_0 /3} \ . 
\eeq

On a compact manifold  the Bianchi identity for the five-form flux leads to the tadpole cancelation condition forcing the total D3-charge of the solution to be zero.
\beq \label{tadpole1}
MK+M^I K_I - M_I K^I + Q^\mathrm{loc}_3= 0 \  , \
\eeq
where the charge of localized D3-brane and O3-plane sources is\footnote{There can be also an additional contribution to $Q^\mathrm{loc}_3$ coming from D7-branes and O7-branes.}
\beq
Q^\mathrm{loc}_3 = N_{D3}-\frac14 N_{O3} \, . \label{tadpole2}
\eeq
The tadpole condition (\ref{tadpole1}) leads to a upper bound on the product of fluxes allowing the cancelation of $C_4$ flux
\begin{equation}
MK \leq | Q^\mathrm{loc}_3|  \,  . \label{tadpole3}
\end{equation}

Recently it was shown \cite{Bena:2018fqc, Blumenhagen:2019qcg} that in the strongly warped region of the Klebanov-Strassler compactification, the light
complex structure field $S$ can be destabilized by the $\overline{D3}$ uplift.  
The potential for the complex structure modulus $S$ involves the fluxes $M$ and $K$, while it depends on the other fluxes only indirectly through the axion-dilaton $\tau$, whose vev is determined by all fluxes. Furthermore, unlike the other ``bulk" moduli, the potential for $S$ is highly affected by the warp factor.
Its functional form, in the Einstein frame, derived in \cite{Douglas:2007tu, Douglas:2008jx} is 
\begin{eqnarray}\label{eq:VKS0}
&& V_{KS} = \frac{\pi^{3/2}}{\kappa_{10}} \frac{g_s}{(\rho + \bar \rho)^3}\left[c \log\frac{\Lambda_0^3}{\left|S\right|} + c'\frac{{g_s (\alpha' M)^2}}{\left|S\right|^{4/3}}\right]^{-1} \left|\frac{M}{2 \pi i} \log\frac{\Lambda_0^3}{S} + i \frac{K}{g_s} \right|^2  \nonumber \\
&& \qquad \simeq \frac{\pi^{3/2} |S|^{4/3}}{\kappa_{10} c' (\alpha' M)^2 (\rho + \bar \rho)^3 }  \left|\frac{M}{2 \pi i} \log\frac{\Lambda_0^3}{S} + i \frac{K}{g_s} \right|^2 \,,
\end{eqnarray}
where in the last line we used the approximation of strong warping. Moreover,  $g_s$ is the stabilized vev of the dilaton, $\rho + \bar \rho=({\rm Vol}_6)^{3/2}$,  
\(c\) denotes the constant value of the warp factor at the UV and will not be relevant here, whereas the constant \(c'\), multiplying the term coming solely from the warp factor, denotes an order one coefficient, whose approximate numerical value was determined in \cite{Douglas:2007tu} to be $ c' \approx 1.18 $.  The potential  (\ref{eq:VKS0}) has a supersymmetric minimum
\begin{equation}
S_{KS} \ =  \ {\Lambda}_0^3 \ e^{- \frac{2 \pi {K}}{ g_s M}} \ , \label{susyKS}
\end{equation}
which is exponentially small for appropriate values of the fluxes $(M,K)$. Since the field $S$ has mass dimension $-3$, whereas the corresponding gauge theory condensate $Z$ has dimension $3$, one can 
write the potential in terms of $Z$ in the following way.  Writing the 10d metric in the form
\begin{equation}
ds^2 = e^{2A} ds_4^2 + e^{-2 A} t^{1/2} ds_6^2 \ , \label{z1}
\end{equation}
where the volume of the internal space is parametrized in terms of $t = \rho + \bar \rho$.  The relation between the 10d and the 4d Newton constant is
\begin{equation}
\frac{1}{\kappa_4^2} =  \frac{V_w}{\kappa_{10}^2} \,,\qquad {\rm where }\qquad V_w = \int d^6 y \sqrt{g_6} e^{-4A}  \ , \label{z2}
\end{equation}
where $V_w$ is a fiducial volume. Using the relation $2  \kappa_{10}^2 = (2 \pi)^7 \alpha'^4$ and redefining the $S$ field according to
\begin{equation}  
S =  \left(2^{3/4} \pi^{1/2} \alpha'\right)^3 Z  \ , \label{z3}
\end{equation}
one arrives at the 4d scalar potential in the Einstein frame\footnote{We take also into account the change of power $1/(\rho+ \bar \rho)^3 \to 1/(\rho+ \bar \rho)^2$ due to the warping,
as argued for in \cite{kklmmt}.}
\begin{equation}\label{eq:VKS}
V_{KS} = \frac{\left|Z\right|^{4/3}} {c' M^2 (\rho + \bar \rho)^2}  \left|\frac{M}{2 \pi i} \log\frac{\Lambda^3}{Z} + i \frac{K}{g_s} \right|^2  \,,
\end{equation}
where we have redefined $\Lambda_0 \rightarrow \Lambda$ analogously to \eqref{z3} and from now on one sets $\kappa_4=1$. The scale $\Lambda$ in (\ref{eq:VKS}) has now mass dimension one. Later on one will define a more canonical dimension-one field $Z \sim Y^3$.

An anti-D3 brane at the tip of the throat uplifts the KS potential \eqref{eq:VKS}. The contribution to the potential is determined from
\begin{equation}\begin{aligned}\label{eq:D3action}
S_{D3} = S_{DBI} + S_{CS}
= - T_3 \int \dd^4 x \sqrt{-g_4} \bigl[1 + \cO(\alpha'^2)\bigr] \pm T_3 \int C_4 \,,
\end{aligned}\end{equation}
where the sign in front of the second term is determined by the charge of the brane, and 
 \(T_3\) is given by  $T_3 = \frac{1}{(2 \pi)^3 \alpha'^2}$.  
For the D3-brane in a background given by \eqref{eq:warpedbackground}, the DBI and the CS pieces of the action cancel each other. Hence, for the \(\overline{\mathrm{D3}}\)-brane they add up and one finds
\begin{equation}
V_{\overline{D3}} = - 2 T_3 C_4 = \frac{2}{(2 \pi)^3 \alpha'^2} H^{-1} \,.
\end{equation}
Using the warp factor given in \eqref{eq:warpfactor} and turning into the 4d Einstein frame, one finally obtains
\begin{equation} \label{VantiD3}
V_{\overline{D3}} = \frac{1}{\pi (\rho + \bar \rho)^2}  \frac{2^{1/3}}{I(\cal T)} \frac{\left|Z\right|^{4/3}}{g_s M^2} \,.
\end{equation}
The \(I(\cal T)\), defined in \eqref{eq:I} is a monotonically decreasing function. Therefore, a \(\overline{\mathrm{D3}}\)-brane has minimal energy if it is placed at the tip of the throat. 
For later convenience we introduce a constant
$ c'' = \frac{2^{1/3}}{I(0)} \approx 1.75 $. 

With these notations and in the highly warped region, the total potential takes the form
\begin{equation}
V_{\rm KS+uplift} = \frac{|Z|^{4/3}  }{(\rho + \bar \rho)^2 c' M^2} \left\{  \left|\frac{M}{2 \pi i} \log\frac{\Lambda^3}{Z} + i \frac{K}{g_s} \right|^2  
+ \frac{c' c''}{\pi g_s}  \right\}  \ . \label{KS+uplift}
\end{equation}
The minimum of the potential with the uplift can be found analytically to be given by \cite{Bena:2018fqc}
\begin{equation}
Z_0 = e^{- \frac{3}{4} \left(1-\sqrt{1- \frac{64 \pi c' c''}{9g_s {M}^2}} \right)} \ {\Lambda}^3 \ e^{- \frac{2 \pi {K}}{ g_s M}}  \ ,  \label{KS+uplift2}
\end{equation}
which clearly displays the disappearence of the non-trivial minimum for small values of the fluxes $g_s M^2 < (64 \pi c'c'')/9$, leading to the condition (\ref{intro1}). 

Some caution is however required in using the scalar potential  (\ref{KS+uplift}).\footnote{We thank A.~Hebecker and L.~Martucci for useful discussions on this point.} In the KS solution
 the $S$ field is not a modulus, but is fixed to its supersymmetric value (\ref{susyKS}). Replacing the warp factor with its $S$-dependence as in (\ref{VantiD3}) is well justified only at the
 minimum and might not be trusted far from it. So the potential  (\ref{KS+uplift}) might not be really trusted ``off-shell'' far from its KS minimum. For large values of $M$-flux, the shift in the vev
 induced by the antibrane uplift is small and the potential should be reliable close to the new minimum. 
 For smaller values of the flux, below to the destabilization point (\ref{intro1}), the runaway behavior depends crucially on the form of the potential for small values of $S$, far away from the old supersymmetric minimum.
Therefore, it cannot be excluded that there are significant corrections to the potential which modify or invalidate the bound \eqref{intro1}.
On the other hand, the results of \cite{Bena:2019sxm} provide further evidence for the validity of \eqref{intro1} and thus also of the potential \eqref{KS+uplift}.
There, a similar bound was obtained from the numerical construction of Klebanov-Strassler black holes \cite{Buchel:2018bzp} which goes beyond the probe approximation and takes the full back reaction on the geometry into account.
Moreover, \cite{Randall:2019ent} compared the analysis of \cite{Bena:2018fqc, Blumenhagen:2019qcg} with the radion stabilization in RS scenarios by the Goldberger-Wise formalism and found qualitative agreement.

In this paper we will not try to resolve these issues and will take the validity of the potential \eqref{KS+uplift} as a working assumption.
Moreover, we mostly work in the regime of large enough $g_s M^2$ where the shift in $S$ induced by the $\overline{D3}$ brane is small.
It is therefore reasonable to assume that in the vicinity of the $\overline{D3}$ minimum possible corrections to the potential are subleading.
Notice, that even in this regime of parameters the effects of the warp factor on the potential \cite{Douglas:2007tu, Douglas:2008jx} are crucial for our following analysis.
Neglecting them corresponds to setting $c' = 0$ and results in a significantly different mass spectrum.


\section{Four-dimensional supergravity description and moduli stabilization}\label{sec:4D}

Adding the nonperturbative term generated by stringy instantons or gaugino condensation amounts to adding to the perturbative superpotential
the KKLT-like terms $W_{KKLT} = W_0+  A e^{- a \rho}$.  The KKLT sector by itself will be defined by
\begin{equation}
{\cal K} = - 3 \log (\rho + \bar \rho) \quad , \quad  W_{KKLT} = W_0+  A e^{- a \rho}   \ , \label{kklt1}
\end{equation}
which leads to the KKLT potential 
\begin{equation}
V_{KKLT} \ = \ \frac{1}{(\rho + \bar \rho)^3}  \left\{  \frac{(\rho + \bar \rho)^2}{3} \left|\partial_\rho W_{KKLT} -  \frac{3}{\rho + \bar \rho} W_{\rm KKLT}\right|^2 -   3 |W_{KKLT}|^2  \right\} \ , \
\label{kklt1}
\end{equation}
which has an AdS minimum. We will show that, under some mild assumptions, the KKLT sector does not affect the conifold destabilization mechanism we found. 
Before showing this, we remind the reader that the uplift can also be described in a manifestly supersymmetric formalism using nonlinear supersymmetry with a nilpotent goldstino 
superfield \cite{nilpotent1, nilpotent2,kallosh}.  
Therefore, if there is a mass gap we should be able to describe the whole action in terms of a supergravity action. Introducing a nilpotent superfield $X$, it is indeed possible to write the K\"ahler potential at the perturbative level as

\begin{eqnarray}
&& {\cal K} = - 3 \log \left( \rho + \bar \rho -  \frac{|X|^2}{3} - \frac{\xi |Z|^{2/3}}{3} \right) -\log\left(- i (\tau - \bar \tau)\right) +  c |Z|^2 \left( \log  \frac{\Lambda^3}{|Z|} + 1 \right) \ ,  \nonumber \\
&& W = W_0 + A e^{- a \rho} + \frac{M}{2 \pi i} Z \left(\log \frac{\Lambda^3}{Z} + 1\right) + K \tau Z + \frac{1}{M} \sqrt{\frac{c''}{\pi}} Z^{2/3} \tau X  \ , \label{4d1}
\end{eqnarray}
where we introduced the (flux-dependent) constant $\xi =  9 c' g_s M^2$. $W_0$ denotes the vev of the bulk or ``UV'' superpotential. 
In what follows, we assume the dilaton to be stabilized $\tau = \frac{i}{g_s}$.
 The field $X$ satisfies the nilpotent constraint $X^2 = 0$. It contains the goldstino $G$ localized on the antibrane and generates the Volkov-Akulov
nonlinear supersymmetric Lagrangian.  The solution of the constraint, in superspace language is
\begin{equation}
X = \frac{GG}{2 F_X} + \sqrt{2} \theta G + F_X \theta^2 \ , \label{4d2}
\end{equation}
where $\theta$ is the fermionic superspace coordinate.  
The simplest  recent string theory examples of Volkov-Akulov nonlinear supersymmetric actions consists of putting a stuck $\overline{D3}$ antibrane on top of an $O3_{-}$ plane \cite{uranga}, which reduces the (anti)brane localized degrees of freedom to only the goldstino \cite{kallosh}. Similar construction, much like the original string vacua with ``brane 
supersymmetry breaking"  \cite{bsb} also generate a nonlinear realization of supersymmetry on the antibranes, as shown explicitly in \cite{dm}. 
The nilpotent constraint eliminates the scalar partner of the goldstino, keeping the auxiliary field $F_X$. Consequently, the scalar potential  is computed from the usual supergravity potential, by setting at the end $X = 0$. The last term in the superpotential reproduces the   antibrane uplift, redshifted by the
S-dependent prefactor.  
Note that the nilpotent goldstino formalism is valid as long as $F_X \not=0$. In the example we consider
(\ref{4d3}), we find\footnote{The factor i can be removed by a redefinition of the field $X$.} that $F_X =  \frac{i}{g_s M} \sqrt{\frac{c^{''}}{\pi}} Z^{2/3}   $ and, since $\langle Z \rangle \not=0$,  the formalism is indeed valid. The stronger the warping the smaller the supersymmetry
breaking. We expect in principle a maximum value of the warping also  from the requirement that states decoupled by the supersymmetry breaking to be heavy enough.  

This K\"ahler potential should be understood as an $Z$-expansion of the general K\"ahler potential derived in \cite{DeWolfe:2002nn} (see also \cite{warpedcomp}), reproducing the metric $G_{S \bar S}$ of \cite{Douglas:2007tu, Douglas:2008jx}.

A naive integration-out of $Z$ would produce an effective constant
\begin{equation} 
W_{0,eff} = W_0 + \frac{M}{2 \pi i} \left( 1+ \sqrt{1-\frac{64 \pi c' c''}{9g_s M^2}}  \right) Z_0 \ . \label{4d3}
\end{equation}

It is convenient in what follows to work with a dimension 1-field $Y$ instead of the dimension $3$ one $Z$. We introduce the convenient definitions 
\begin{eqnarray}
&& Y = \left(\frac{\xi}{3}\right)^{1/2} Z^{1/3}  =  \left(\frac{3c'g_s M^2}{2^{3/2} \pi \alpha'^2}\right)^{1/2} S^{1/3} \ ,  \nonumber \\
&& {\ \tilde M} = \left(\frac{3 }{\xi}\right)^{3/2} M \ , \quad  {\ \tilde K} = \left(\frac{3 }{\xi}\right)^{3/2} K \ ,  \quad {\tilde \Lambda}_0 = \left(\frac{\xi}{3}\right)^{1/2} \Lambda \ , \quad 
{\tilde c''} = \frac{3 }{\xi g_s M} \sqrt{\frac{c''}{\pi}} \ . \label{4d4}
\end{eqnarray}
With these definitions, freezing the dilaton and in the region of strong warping, where the perturbative K\"ahler potential for S is negligible, the effective action  (\ref{4d1}) becomes

\begin{eqnarray}
&& {\cal K} = - 3 \log \left( \rho + \bar \rho -  \frac{|X|^2}{3} -  |Y|^2   \right)  \ ,  \nonumber \\
&& W = W_0 + A e^{- a \rho} + \frac{\tilde M}{2 \pi i} Y^3 \left(3 \log \frac{\tilde\Lambda_0}{Y} + 1\right) + \frac{i \tilde K}{g_s}  Y^3 + {\tilde c}^{''} Y^2 X  \ . \label{4d5}
\end{eqnarray}
Therefore, the SUGRA scalar potential of the model (\ref{4d5}) can be written in the form 
\begin{equation}
V =  \frac{1}{r^2}  \left\{ 3 |Y|^{4}  \left| \frac{3 \tilde M}{2 \pi i} \log\frac{\tilde \Lambda_0}{Y} + i \frac{\tilde K}{g_s} \right|^2  
+ \left|\tilde{c''} Y^2\right|^2 + \left|\partial_\rho W -  \frac{3}{\rho + \bar \rho} W_{\rm eff}\right|^2 -   \frac{3}{\rho + \bar \rho}  \left|W_{\rm eff}\right|^2  \right\}
  \ , \label{4d6}
\end{equation}
where  we defined
\begin{equation}
W_{\rm eff} = W_0 + A e^{-a \rho} +  \frac{\tilde M}{2 \pi i} Y^3   \quad {\rm and } \quad  r = \rho + \bar \rho -  |Y|^2  \ . \label{4d7}
\end{equation}
The first two terms in  (\ref{4d6}) recover the  potential $V_{\rm KS+uplift}$ displayed in   (\ref{KS+uplift}) coming from the fluxes and the $\overline{D3}$ uplift, written in terms of the dimension-one field $Y$ and in the small $Y$ limit.  The last two terms in the potential  (\ref{4d6}) reduce, for $Y=0$, to the KKLT potential for $\rho$.  

\subsection{Vacuum structure and mass scales}\label{sec:masses}

The scalar potential (\ref{4d6}) has an almost decoupled structure, between the KS and the KKLT sector.  In the case that one of the moduli is significantly heavier than the other, this implies
that the decoupled KS+uplift and the KKLT+uplift minima will be a good zeroth order approximation. The later ones are determined as
\begin{equation}
\left.\frac{\partial V_{\rm KS+uplift}}{\partial Y}\right|_{Y_0} = 0 \quad , \quad \left.\frac{\partial V_{\rm KKLT+uplift}}{\partial \rho}\right|_{\rho_0} = 0   \ . \label{v1}
\end{equation}
The explicit values are
\begin{eqnarray}
&& Y_0 =  e^{- \frac{2 \pi \epsilon_0}{3 \tilde M}}  \ {\tilde \Lambda}_0 \ e^{- \frac{2 \pi {\tilde K}}{3 g_s \tilde M}}  \quad \leftrightarrow  \quad 
Z_0 = e^{- \frac{3}{4} \left(1-\sqrt{1- \frac{64 \pi c' {c}^{''}}{9g_s {M}^2}} \right)} \ {\Lambda}_0^3 \ e^{- \frac{2 \pi {K}}{ g_s M}}  \ , \nonumber \\
&& \left[  a (\rho_0 + {\bar \rho}_0) + 5  \right] A e^{-a \rho_0} = - 3 W_0   \ . \label{v2}
\end{eqnarray}
The constant $\epsilon_0$ above is given by
\begin{equation}
\epsilon_0 = \frac{3 \tilde M}{2 \pi} \log \frac{\tilde \Lambda_0}{Y_0} - \frac{\tilde K}{g_s} = \frac{1}{4} \left( \frac{3 \tilde M}{2 \pi} - \sqrt{\left(\frac{3 \tilde M}{2 \pi}\right)^2 - \frac{48 \tilde c''{}^2}{9 }} \right)
=  \frac{3}{8 \pi M^2 (3 c' g_s)^{3/2} }  \left(1-\sqrt{1- \frac{64 \pi c' c''}{9g_s {M}^2}} \right)
 \ , \label{v3}
\end{equation}
and measures the deviation of the minimum of $S_0$ from the supersymmetric one (\ref{susyKS})  before the uplift.  Notice that the higher the flux $M$ the smaller the deviation of the vev of $Z$ from its
 supersymmetric value. 
 The main result of \cite{Bena:2018fqc} was that the existence of a non-trivial minimum for $S$ implies a minimum value for the $M$ flux
 \begin{equation}
g_s M^2 \ \geq M_{\rm min}^2 = \frac{64 \pi c' c''}{9} \simeq (6.8)^2    \ . \label{v4}
\end{equation}
 The cancelation of the cosmological constant after uplift translates into the tuning
\begin{equation} 
 \left(  |{\tilde c''}|^2 + 3 \epsilon_0^2 \right) |Y_0|^4 \simeq \frac{3 |W_0|^2}{\rho_0 + \bar \rho_0}  \quad \to \quad W_0 \simeq \epsilon_0 (\rho_0+\bar \rho_0)^{1/2} Y_0^2 \ ,   \label{v5}
 \end{equation}
 where in the last estimate we neglected $ |{\tilde c}| \ll \epsilon_0$, valid for not extremely large values of the flux. Notice that an uplift to zero cosmological constant requires $W_0 \sim Z_0^{2/3}$ which 
 is much larger (in the strong warping regime) than the last term in  (\ref{4d3}), induced by integrating-out the field $Z$.  The mass of the moduli fields are then readily computed. We denote by
 $m_{Y,\pm}$ the two mass eigenstates  of the complex field $Y$. At the leading order 
 one finds
\begin{eqnarray}
&& m_{Y,+}^2 = \frac{\tilde M}{4 \pi} \frac{Y_0^2}{\rho_0+{\bar \rho}_0} =  \frac{Z_0^{2/3}}{8 \pi \rho_0 (3c'g_s)^{1/2}} \quad , \quad m_{Y,-}^2 =  m_{Y,+}^2
\sqrt{1- \frac{M_{\rm min}^2}{g_s {M}^2}}
\ ,  \nonumber \\
&& m_{\rho {\bar \rho}}^2 =   \frac{a^2 W_0^2}{\rho_0 + \bar \rho_0}  \quad , \quad m_{\rho \rho}^2 \ll m_{\rho {\bar \rho}}^2   \ .  \label{v6}
\end{eqnarray}
After imposing the cancelation of the vacuum energy, the ratio of the moduli masses is given by
\begin{equation}
\frac{m_\rho^2}{m_Y^2} \sim  \left(1-\sqrt{1- \frac{M_{\rm min}^2}{g_s {M}^2}} \right)^2   \frac{ a^2  
 (3 c')^{5/2}  \rho_0 Z_0^{2/3} }{32 \pi^2 g_s^{1/2} }   \ .  \label{v7}
\end{equation}
The $Z$ modulus is therefore heavier than $\rho$ provided that 
\begin{equation}
Z_0 \ll  \left[  \frac{32 \pi^2 g_s^{1/2}}{\left(1-\sqrt{1- \frac{M_{\rm min}^2}{g_s {M}^2}} \right)^2 (3 c')^{5/2}} \right]^{\frac{3}{2}}   \frac{1}{(a^2 \rho_0)^{3/2}}  \ .  \label{v8}
\end{equation}
This is also the condition of validity for the (quasi) decoupling of the  KS and KKLT sectors, that was our starting assumption in finding the vacuum structure. 
One can aposteriori check that the shifts in the minima after coupling the KS+uplift and the KKLT sector are parametrically $\delta Y/Y_0 \sim \delta \rho/\rho_0 \sim  Y_0 \sqrt{\rho_0}$, which
are small precisely when  (\ref{v8}) is fulfilled. 
 The constraint (\ref{v8}) is easy to satisfy for
large fluxes. The strongest constraint arises for small fluxes. For example, close to the critical value (\ref{v4}) and in the case of stringy instantons $a = 2 \pi$,
 (\ref{v8}) gives roughly $Z_0 \ll g_s^{3/4}/(3\rho_0)^{3/2}$. 
 Since
\begin{equation}
\frac{2 \pi K}{g_s M} = \frac{2 \pi M K}{g_s M^2 }  \leq \frac{2 \pi |Q^\mathrm{loc}_3| }{(6.8)^2}  \  ,  \label{v9}
\end{equation} 
a  given value of the warp factor translates into  a minimum value of the localized D3 charge $Q^\mathrm{loc}_3$, according to 
\begin{equation}
\left|Q^\mathrm{loc}_3\right| \geq \frac{(6.8)^2}{2 \pi} \left| \log Z_0 \right|  \,.  \label{v10}
\end{equation}

For example, typical KKLT values $\rho_0 \sim 20-50$ and for $\Lambda_0 \sim M_P$ imply  $Z_0 < 10^{-2}$ to satisfy (\ref{v8}),which would imply $ |Q^\mathrm{loc}_3| \geq 40-50$. As another example, the gravitino mass is given by
\begin{equation}
m_{3/2} \simeq \frac{W_0}{(\rho_0+{\bar \rho}_0)^{3/2}} \sim \frac{\epsilon_0 Y_0^2}{\rho_0} \  ,  \label{v11}
\end{equation}
where the last estimate is order of magnitude only. 
 TeV values of the gravitino mass $m_{3/2} \sim $ TeV would require  $Z_0 \sim 10^{-21}$, which translates into $ |Q^\mathrm{loc}_3| \geq 350-400$.  Whereas this is possible in F-theory, it is challenging to construct
explicit examples with such large localized D3 charges \cite{blaback}. 
\subsection{Contribution to supersymmetry breaking}

The contribution of various fields to supersymmetry breaking is encoded in the auxiliary fields, in terms of which one can write the scalar potential as
\begin{equation}
V = {\cal K}_{I \bar J} F^I F^{\bar J} - 3 m_{3/2}^2 \,, \qquad {\rm where } \qquad F^I = e^{{\cal K}/2} {\cal K}^{I \bar J} \overline{D_J W}  \  .  \label{sb1}
\end{equation} 
Using the vev's obtained in the previous section, one finds
 \begin{equation}
 D_X W = \frac{1}{g_sM} \sqrt{\frac{c^{''}}{\pi}} Z_0^{2/3} \,, \qquad D_Y W \simeq - 3 i \epsilon_0 Y_0^2 \,, \qquad D_\rho W \simeq - \frac{6 W_0}{a (\rho_0+\bar \rho_0)^2}  \  .  \label{sb2}
  \end{equation}
 Even if the fields $Y$ and $\rho$ mix in the K\"ahler potential, it can be shown that it is a good approximation to neglect the mixing and to define individual contributions
 to SUSY breaking, according to
 \begin{equation} 
f_I \equiv e^{{\cal K}/2}   ({\cal K}^{I \bar I})^{1/2} \overline{D_I W} \,, \qquad V \simeq \sum_I |f_I|^2 - 3 m_{3/2}^2  \  .  \label{sb3}
\end{equation}
Then (\ref{sb2}) leads to
\begin{equation}
f_X \sim f_Y \sim m_{3/2} \,, \qquad f_\rho \sim \frac{m_{3/2}}{a (\rho_0+\bar \rho_0)}  \  .  \label{sb4}
\end{equation}
It is interesting that the conifold field $Y \sim Z^{1/3}$ has a large contribution to supersymmetry breaking, at the same order as the one of the uplift (nilpotent) field $X$. Both contributions 
are localized at the tip of the throat. Notice that a contribution to supersymmetry breaking of the complex structure field $Z$ could be interpreted as an effective generation of an $(1,2)$ flux. 
On the other hand,  the contribution to supersymmetry breaking of the K\"ahler modulus $\rho$, which propagates across the whole bulk, is suppressed by a factor
of order $1/(a \rho_0)$. This will have consequences on the transmission of supersymmetry breaking into the matter sector.  
\subsection{Comments on the dS swampland conjecture}

It was recently conjectured \cite{obied} that in all controlled compactifications 
\begin{equation}
| \nabla V | \geq a V \qquad {\rm or} \qquad {\rm min}  (\nabla_i \nabla_j V) \leq -a' V \ , \label{sw1}
\end{equation}
where $a,a' >0 $ are $\mathcal{O}(1)$ numbers.  The second condition is not constraining in our case, therefore we discuss the first one. Since the KS and the KKLT sector are approximately decoupled 
in the regime $m_\rho \ll m_Y$, we can concentrate on the KS plus the uplift sector, assuming the K\"ahler modulus $\rho$ is stabilized, described by
\begin{equation}
{\cal L}_Y = \frac{3}{\rho+\bar \rho} |\partial Y|^2 - V_{\rm KS+uplift}  \ , \label{sw2}
\end{equation}
where the appropriate scalar potential is given in (\ref{KS+uplift}).  We therefore compute
 \begin{equation}
\frac{\left|\nabla_Y V\right|}{V} = \frac{\sqrt{G^{Y\bar Y} \partial_Y V \partial_{\bar Y} V}}{V} \,.  \label{sw3}
\end{equation} 
If $\sqrt{g_s} M \geq M_{\rm min}$, there is a dS minimum and the dS conjecture is violated. If this is realized in string theory depends on the existence of compactifications with
large localized D3 charges \cite{blaback}. 
Following the same steps and arguments as in  \cite{Bena:2018fqc}, another check can be performed for small flux $\sqrt{g_s} M < M_{min}$, where the dS minimum disappear.
One finds
 \begin{equation}
\frac{\left|\nabla_Y V\right|}{V} \geq \frac{2}{\left| Y \right|} \sqrt{\frac{\rho+\bar \rho}{3}} \left(1 - \frac{ \sqrt{g_s}M}{M_\text{min}}\right) \,  \ . \label{sw4}
\end{equation}
This is generically satisfied in the limit of strong warping. It is however amusing to notice that  by imposing $a \sim 1$ one obtains a condition  parametrically of the type (\ref{v8}), although the two
conditions apply to different cases.  

On the other hand, a sufficient (but not necessary) condition to satisfy the dS conjecture would be to select large enough values of the $(0,3)$ flux parameter $W_0$  to forbid
an uplift to positive vacuum energy. Using the results and notations from Section~\ref{sec:masses}, this condition reads
\begin{equation}
W_0 \geq \epsilon_0 (\rho_0 + \bar \rho_0)^{1/2} Y_0^2 \ . \label{sw5}
\end{equation}
 
\section{Adding matter: soft terms}
\label{sec:softterms}

The Klebanov-Strassler throat  can generate an exponential hierarchy for the scale of supersymmetry breaking in the observable sector. 
The matter fields $Q_i$ are defined as usual by the properties  $\langle Q_i \rangle =0 $, $F^i = 0$. 
 The 4d supergravity lagrangian contains hidden sector (moduli) fields called $T_\alpha$ in what follows ($T_\alpha= X,Y,\rho$ in our case), coupled to the matter fields $Q_i$. 
 The K\"ahler potential and superpotential are defined by an expansion in powers of the matter fields
 \begin{equation}\begin{aligned}
 {\cal K} &= {\hat {\cal K}} (T_\alpha, {\overline T}_{\alpha}) + K_{i \jbar} (T_\alpha, {\overline T}_{\alpha}) Q^i {\overline Q}^{\jbar} + \frac{1}{2} \left( Z_{i j} (T_\alpha, {\overline T}_{\alpha}) Q^i {Q}^{ j} + {\rm h.c.} \right) 
 \,, \\
 W &= {\hat W} (T_\alpha) + \frac{1}{2}  {\tilde \mu}_{i j} (T_\alpha) Q^i {Q}^{ j} +  \frac{1}{3}  {\tilde Y}_{i jk} (T_\alpha) Q^i {Q}^{ j} Q^k + \cdots    \,.   \label{soft1}
\end{aligned}\end{equation}
The low-energy softly broken supersymmetric lagrangian is defined by the superpotential and soft scalar potential
\begin{equation}\begin{aligned}    
W_{\rm eff} &= \frac{1}{2}  {\mu}_{i j} Q^i {Q}^{ j} +  \frac{1}{3}  {Y}_{i jk}  Q^i {Q}^{ j} Q^k \,, \\ 
  {\cal L}_{\rm soft} &= - m_{i \jbar}^2 q^i q^{\jbar} - \left( \frac{1}{2}  {B}_{i j} q^i {q}^{ j} +  \frac{1}{3}  {A}_{i jk}  q^i {q}^{ j} q^k + \frac{1}{2} M_a \lambda_a \lambda_a + {\rm h.c.} \right) \,,   \    \label{soft2}
\end{aligned}\end{equation} 
where $Y_{ijk} = e^{K/2}  {\tilde Y}_{i jk}$. 
For zero cosmological constant, the general  tree-level expressions
for soft terms and supersymmetric $\mu$-terms in 4d supergravity,  are given by \cite{kl}
\begin{equation}\begin{aligned}
m_{i \jbar}^2 &= m_{3/2}^2 (G_{i \jbar} - G^{\alpha} G^{ \bar \beta} R_{i {\jbar} \alpha \bar \beta} )  =  m_{3/2}^2 K_{i \jbar} - F^{\alpha}  F^{\bar \beta} R_{i \jbar \alpha \bar \beta}  \ ,  \\
 \mu_{ij} &= m_{3/2} \nabla_i G_j  \ = \ e^{\frac{K}{2}}  {\tilde \mu}_{i j}  + m_{3/2} Z_{ij} -  F^{\bar \alpha} \partial_{\bar \alpha}   Z_{ij} \ , \\
A_{i jk} &= m_{3/2}^2 (  3 \nabla_i G_j +  G^{\alpha}  \nabla_i \nabla_j \nabla_k G_\alpha )   = 
 (m_{3/2} - F^{\alpha} \partial_\alpha \log m_{3/2} ) Y_{ijk} + F^{\alpha} \partial_\alpha Y_{ijk} - 3 F^{\alpha}  \Gamma_{\alpha (i}^l Y_{ljk)}  \  ,   \\
 B_{ij} &= m_{3/2}^2 (  2 \nabla_i G_j +  G^{\alpha}  \nabla_i \nabla_j G_\alpha )  = 2 m_{3/2}^2 Z_{ij} - m_{3/2} F^{\bar \alpha} \partial_{\bar \alpha} Z_{ij} +   \\
 &\qquad m_{3/2} F^{\alpha} (\partial_\alpha Z_{ij} - \Gamma_{\alpha i}^k Z_{kj}- \Gamma_{\alpha j}^k Z_{ki}) - F^{\alpha} F^{\bar \beta} 
 (Z_{ij \alpha \bar \beta} - \Gamma_{\alpha i}^k Z_{kj \bar \beta}- \Gamma_{\alpha j}^k Z_{ki \bar \beta})   \\
&\qquad - e^{\frac{\cal K}{2}} {\tilde \mu}_{ij} m_{3/2} + e^{\frac{\cal K}{2}}  F^{\alpha} (\partial_\alpha {\tilde \mu}_{ij} + \frac{1}{2} {\hat K}_{\alpha} {\tilde \mu}_{ij} - \Gamma_{\alpha i}^l {\tilde \mu}_{lj} 
- \Gamma_{\alpha j}^l {\tilde \mu}_{il} ) \ ,  \\
 M_{1/2}^a &= \frac{1}{2} g_a^2 F^{\alpha} \partial_\alpha f_a  \ , \label{soft3}
\end{aligned}\end{equation}
where $\alpha ,\beta $ are hidden sector supersymmetry breaking indices, $f_a$ is the gauge kinetic function for the gauge group factor $G_a$ and some basic definitions are given in the Appendix. 
For geometrical separation in the internal space (sequestering)  or in no-scale like models, tree level soft masses are zero or highly suppressed and one-loop contributions
become relevant.  One loop contributions to gaugino masses, called anomaly-mediated contributions in \cite{sequestering1},  are given in general by  \cite{sequestering2}
\begin{equation}
M_{1/2}^a =  - \frac{g_a^2}{16 \pi^2} \left\{  (3 T_G^a - T_R^a) m_{3/2} +  ( T_G^a - T_R^a) K_{\alpha} F^{\alpha} +   \frac{2 T_R^a}{d_R}  (\log \det K_{i {\jbar},R^a})_{\alpha} F^{\alpha}  \right\} \ , 
  \label{soft4}
\end{equation}
where $T_G^a $ is the Dynkin index for the adjoint representation of the gauge group and  $T_G^a $, $K_{i {\bar j},R^a}$ are the Dynkin index for the chiral matter fields in representation $R^a$ and
their K\"ahler metric, respectively. The complete one-loop expression for the other soft terms is more involved \cite{maryk}. In the limit of small hidden-sector vev's, the one-loop induced anomaly mediated
contributions take forms of the type $m_{\rm soft} \sim (b g^2) / (16 \pi^2) m_{3/2}$, depending on beta functions $b$ (and anomalous dimensions) of the low-energy spectrum.  
 
 On general grounds, for matter fields which are not sequestered from the KS throat, one expects soft masses of order $m_{3/2}$. For fields far away from the throat, from the four-dimensional
 perspective, one could anticipate soft terms generated by the K\"ahler modulus contribution to supersymmetry breaking $F^{\rho}$ and one-loop anomaly-mediated contributions. 
 These two contributions are  similar in size, as emphasized in  \cite{Choi:2005ge} and parametrically (one-loop) suppressed compared to the gravitino mass.
  
\subsection{Distant D3 "matter" branes}

One test of the action (\ref{4d1}) is to add distant, from the throat,  D3 ``matter'' branes.  
Denoting by $Q_i$ the (distant) D3 brane superfields, the tree-level action in type IIB orientifolds including them was derived in \cite{ggjl}. In the absence of warping,
the  K\"ahler and the complex structure moduli spaces do not talk to each other, except for the holomorphic terms of coefficients called $z_{ij}$ below. The new ingredient from the warping is that the main
contribution to the complex structure field  $Z$ K\"ahler potential changes drastically (in the absence of matter fields) according to the previous Sections, whereas the perturbative contribution 
is negligible.  Taking this into account, we arrive at the following effective action
\begin{equation}\begin{aligned}
{\cal K} &= - 3 \log \left( \rho + \bar \rho -  \frac{|X|^2}{3} - \frac{\xi |Z|^{2/3}}{3} - |Q_i|^2  - z'_{ij} ( {\bar Z} Q_i Q_j + {\rm h.c.}  )  \right)   \\
&=  - 3 \log \left( \rho + \bar \rho -  \frac{|X|^2}{3} - |Y|^{2} - |Q_i|^2  - z_{ij} ( {\bar Y}^3 Q_i Q_j + {\rm h.c.}  ) \right) \equiv - 3 \log r  \ ,   \\
 W &= W_1 (\rho,Y,X) + W_2 (Q_i)  \ .  \label{m2}
\end{aligned}\end{equation}
Notice that (\ref{m2}) has a sequestered form, except the terms in $z_{ij}$. 
Using the Appendix, it can be shown that in this case  the Riemann tensor satisfies the identity
\begin{equation}
R_{i {\jbar} \alpha \bar \beta}  =  \frac{1}{3}  G_{i \jbar}  G_{\alpha \bar \beta}    \  ,  \label{m3}
\end{equation}
where $\alpha,\beta=X,Z,\rho$ are moduli indices and $i,j$ are matter indices, such that the tree-level scalar soft masses $m_{i \jbar}^2$ in  (\ref{soft3}) vanish after reinforcing the cancelation of the cosmological constant $G^\alpha G_\alpha = 3$. The tree-level A-terms can be written more explicitly
as
\begin{equation}
  A_{i jk} = m_{3/2}^2 \left[  3 G_{ijk} +  G^{\alpha}  ( \partial_\alpha G_{ijk} - \Gamma_{i \alpha}^m G_{jkm}  - \Gamma_{j \alpha}^m G_{ikm} - \Gamma_{k \alpha}^m G_{ijm})  \right]       \   \label{m4}
\end{equation}
and they turn out to also vanish for zero cosmological constant. 

The tree-level supersymmetric masses in our case are given by
\begin{equation}
\mu_{ij} = m_{3/2} (\partial_i G_j - \Gamma_{i j}^\alpha G_{\alpha}) = - 2 z_{ij} {\bar y}^2  (2 \bar y W_T + 3 W_y) e^{\frac{\cal K}{2}} \sim \mathcal{O}\left(\frac{Y_0^4}{\rho_0^{3/2}}\right)   \sim 
\mathcal{O}\left(\frac{Z_0^{4/3}}{\rho_0^{3/2}}\right)  \ .   \label{m5}
\end{equation}
This result is in qualitative agreement with the action obtained from the compactification worked out in \cite{ggjl}. It is however important to remember that our effective action is only valid in the leading order in an expansion in powers of $Z^{2/3}$. The leading order contribution to the $\mu$-terms is small
and subleading and probably a mass of order $Z^{4/3}$ is quantitatively not fully  under control.  
  The $B \mu$-like  terms $B_{i j}$ in \eqref{soft3} are of order $\mathcal{O}(Y^6) \sim \mathcal{O}(Z^2)$, but probably their computation is also not fully reliable. It is also possible that the
  coefficients $z_{ij}$ are warped down such that the real values of $\mu_{ij},B_{ij}$ is even smaller than the one we estimated here. 

The tree-level action of the distant D3 branes seem to realize therefore, at the leading  order in a power expansion of $S^{2/3}$, an approximate sequestered case \cite{sequestering1} where all tree-level ``soft terms'' for matter  D3 fields are zero (or very small). However, it was argued in
\cite{Berg:2010ha}  that sequestering is broken by the coupling of D3 branes to the gaugino condensates on D7 branes or on Euclidian D3 branes providing the nonperturbative superpotential
for the K\"ahler modulus $\rho$.  This effect, coming from threshold corrections, breaks the sequestered form of the superpotential (\ref{m2}), by generating terms
of the type $w (Q) e^{-a \rho}$. Such terms generate soft terms, but not non-holomorphic soft scalar masses. On the other hand, it was shown in \cite{Choi:2005ge,Berg:2010ha} that small values of the latter,
suppressed with respect to the gravitino mass, are also generated by the antibrane uplift. In our manifestly supersymmetric nilpotent formalism, such terms can be generated by breaking sequestering in the 
K\"ahler potential.  Breaking sequestering and agreement with previous results require therefore a modification of our effective action   (\ref{m2}) by
\be
\Delta K = \frac{D |X|^2 {\bar Q}_i Q_i}{r^4} \quad , \quad \Delta W = w(Q_i) e^{-a \rho} \ ,  \label{m05}
\ee
where $D$ is a constant and $w(Q_i)$ a holomorphic gauge invariant function of matter chiral fields. The form of $\Delta K$, which should be treated as a small perturbation of the K\"ahler potential in (\ref{m2}), is determined by matching the resulting generated non-holomorphic scalar masses to previous results,  $m_0^2 \sim m_{3/2}^2/(ar)^2$ for the canonically normalized  matter field \cite{Choi:2005ge,Berg:2010ha}. The resulting soft terms are suppressed
and similar in size to the anomaly mediated terms discussed below.  It would be interesting to understand from first principles the term  $\Delta K$ breaking  sequestering. 

Notice that the $\overline{D3}$ brane interacts via loops of open strings, or equivalently, by tree-level exchange of closed strings, with distant D3 branes.  This is a higher-dimensional  tiny  branes-antibrane interaction which is actually a one-loop effect, that has to be added to the effective action above. 
The sequestering is approximative therefore from several viewpoints: the threshold effects on D7 branes (or Euclidian D3 branes) and the antibrane-D3 brane distant interactions. In addition, there are small distant branes mass terms related to the contribution to supersymmetry breaking of the conifold field $Z$. They could maybe have an interpretation
 since $F^Z\not=0$ has an effect similar to the generation of an effective $(1,2)$ flux. 

From  a four-dimensional perspective, one can also contemplate  calculating the one-loop anomaly-mediated type contributions. For the effective action (\ref{m2}), 
the one-loop gaugino masses  reduce at leading order to the universal term
\begin{equation}
M_{1/2}^a =  - \frac{g_a^2}{16 \pi^2}  (3 T_G^a - T_R^a) (m_{3/2} +  \frac{1}{3} K_{\alpha} F^{\alpha} )  \simeq  \  - \frac{g_a^2}{16 \pi^2}  (3 T_G^a - T_R^a) m_{3/2} \   . 
  \label{ano1}
\end{equation}

 \subsection{D3 branes-antibrane interactions: a higher-dimensional perspective}
 
In addition to the threshold effects on D7 branes or euclidian D3 branes, the sequestering is broken explicitly by the distant branes-antibrane interactions, which generate an interaction between the antibrane and the distant antibranes \cite{kklmmt}.  This interaction can be 
 described by a correction to the K\"ahler potential, breaking the sequestered structure,  of the type
 \begin{equation}
\Delta {\cal K} = \frac{\xi'_z |Z|^{4/3} |X|^2}{ r ({\bf r_0}+{\bf Q})^4}  \  ,  \label{m6}
\end{equation}
where $d^2 = {\bf r_0}^2$ is the radial distance between the $D3$ branes and the antibrane. Indeed, for ${\bf Q}_i=0$ this term changes the metric ${\cal K}_{X \bar X} = (1+ \frac{\xi'_z |Z|^{4/3}}{d^4}) \frac{1}{r}$  
and changes the scalar potential generated by the uplift sector to
\begin{equation}
V_{\overline{D3}} + V_{D3-\overline{D3}} = {\cal K}^{X \bar X} e^{\cal K} |D_X W|^2  \simeq \frac{c'' |Z|^{4/3}}{\pi ( \rho + \bar \rho)^2 g_s M^2} \left( 1 - \frac{\xi'_z |Z|^{4/3}}{d^4} \right)  \equiv 2 T_3 H^{-1} (r_0) 
\left[ 1 - \frac{1}{N} \frac{H(r_1)}{H(r_0)}  \right]  \  ,  \label{m7}
\end{equation} 
where $r_0$  ($r_1$) is the radial position of the $\overline{D3}$ brane (matter $D3$) branes and we performed a leading-order expansion in powers of $Z\ll1$. An expansion in powers of the matter fields gives the antibrane-distant D3 branes fields interaction
 \begin{equation}
V_{D3-\overline{D3}}  = - \frac{ \xi'_z c'' |Z|^{8/3} }{ \pi r^2 d^4 g_s M^2} \left[ 1 - \frac{4{\bf r_0} {\bf Q}}{d^2} - \frac{2 {\bf Q}^2}{d^2} + \frac{12 ({\bf r_0} {\bf Q})^2}{d^4} + \cdots  \right]  \  .  \label{m8}
\end{equation}
In the higher-dimensional approach approach, the second derivative of this scalar potential gives the mass matrix for ${\bf Q}_i$, which lead to tachyonic directions \cite{pablo}. It is however useful to cast
 the problem as a correction to the 4d K\"ahler potential. By using the dimension-1 field $Y$, it is given by 
 \begin{equation}
\Delta {\cal K} = \frac{\xi' |Y|^{4} |X|^2}{ r d^4} \left[ 1 - \frac{4{\bf r_0} {\bf Q}}{d^2} - \frac{2 {\bf Q}^2}{d^2} + \frac{12 ({\bf r_0} {\bf Q})^2}{d^4} + \cdots  \right]  \  ,  \label{m9}
\end{equation}
where $\xi' = (3 c' g_s M^2)^{-2} \xi'_z$. 
The linear term signifies that the distant D3 brane does not sit at an extremum, which can be remedied, in case of orientifolds, by adding an image brane at $-{\bf r_0}$. The quadratic terms
are more conveniently written in a complex basis $\Phi_i,z_i$, $i=1,2,3$, by introducing $\Phi_1 = \frac{1}{\sqrt{2}} (Q_1+i Q_2)$,  $z_1 = \frac{1}{\sqrt{2}} (y_1+i y_2)$, etc. One gets
 \begin{equation}
\Delta {\cal K} = \frac{\xi' |Y|^{4} |X|^2}{ 4 r |z|^4} \left[ 1  - \frac{2}{|z|^2} (\delta_{i \jbar} - \frac{3 z_i \bar z_{\jbar}}{4 |z|^2} ) {\bar \Phi}^i \Phi^{\jbar}+ 
\frac{3}{4 |z|^4}  ( \bar z_{\ibar} \bar z_{\jbar} {\Phi}^{\ibar} \Phi^{\jbar}+  {\rm h.c. }) + \cdots  \right]  \  .  \label{m10}
\end{equation}
The non-holomorphic piece  ${\bar \Phi}_i \Phi_j $ changes the metric for matter fields and changes the Riemann tensor of the K\"ahler manifold $R = R^{(0)} + R^{(1)} $. 
It generates an additional contribution to the soft scalar masses
\begin{equation}
m_{i \jbar}^2 = - m_{3/2}^2 |G^X|^2  R_{i \jbar X \bar X}^{(1)} = - |F^X|^2  R_{i \jbar X \bar X}^{(1)} = |F^X|^2  \frac{\xi' |Y|^{4}}{ 2 r |z|^6} (\delta_{i \jbar} - \frac{3 z_i \bar z_{\jbar}}{4 |z|^2} ) \  ,  \label{m11}
\end{equation}
which are positive definite and of order $m_0^2 \sim Y_0^{8}/(\rho_0^2 |z|^6)$, in agreement with the masses computed from the higher-dimensional  potential  (\ref{m8}).  The holomorphic soft masses are 
then given by
\begin{equation}
B_{ij} =  |F^X|^2  \frac{3 \xi' |Y|^{4}}{ 16 r |z|^8}  {z}_i z_j   \  .  \label{m12}
\end{equation}
Notice that they are smaller ($\mathcal{O}(Y^8)$) than the tree-level 4d supergravity computation of the previous section.  They are of the same order than the non-holomorphic masses $m_{i \jbar}^2$ and generate tachyonic masses after diagonalization \cite{pablo}.  However, in a realistic compactification the distant D3 branes have to experience additional projections of the spectrum in order to make it compatible with a MSSM one. For example, it should sit at an orbifold or orientifold singularity.
If this is achieved, the only sector that can have  $B$ terms is the Higgs sector. A Higgs vev could lead to correct electroweak symmetry breaking if its value is in the TeV range, therefore
if the soft terms are in the TeV range. The size of soft terms provided by   (\ref{m11})- (\ref{m12}) can be re-expressed as a function of the gravitino mass and the distance from the D3 antibrane.
Their typical size is then worked out to be of order
$ m_{soft} \ \sim \ \left(\frac{\rho_0}{|z|}\right)^3  \ \frac{m_{3/2}^2}{M_P }$. If the one-loop contributions would be smaller, 
intermediate values of $m_{3/2}$ can lead therefore to TeV values for masses. 
However, as discussed in the previous section, from a four-dimensional perspective, threshold corrections on D7 branes or Euclidian D3 branes and other one-loop corrections 
are expected to break sequestering and generate suppressed soft terms \cite{Choi:2005ge, Berg:2010ha} with respect to the gravitino mass, similar in size to anomaly mediated
contributions  \cite{sequestering1}.  
Their size is typically a one-loop factor times $m_{3/2}$ and will therefore dominate over the (smaller) contributions discussed above. 
It would be interesting to evaluate from a higher-dimensional perspective, based
on anti-brane-distant branes interactions, the one-loop generated soft terms and compare them to the four-dimensional expressions. This is however beyond the scope
of the current letter.

Notice that, if one uses some distant D3 branes for inflation, then a scalar potential of the type (\ref{m7}) becomes the inflationary energy scale. From the current bounds on the tensor to scalar
ratio $r \leq 0.1$, one gets another bound on the acceptable values of $Z_0$
\begin{equation}
V_{\rm inf} = V_{D3-\overline{D3}} \leq 10^{16} \, \mathrm{GeV} \quad \to \quad \frac{Z_0^{4/3}}{\rho_0^2} \leq 10^{-9}   \  .  \label{m14}
\end{equation}
For a KKLT type scenario, this implies $Z_0 \leq 10^{-5}$ and therefore again a largish contribution to the localized D3 flux $|Q_3^{\rm loc}| > \mathcal{O}(80)$, for $\Lambda_0 \sim M_P$. 

\subsection{4d versus higher-dimensional description}

It was shown in \cite{Blumenhagen:2019qcg} that Kaluza-Klein states localized on the KS throat have masses parametrically of the same order as the mass of the field $Z$. 
Then a full 4d description of the dynamics is probably not an accurate approximation, unless one goes to low energies and integrates the whole KK tower and the field $Z$. This it difficult 
to do in practice. On the other hand, the gravitino mass is small enough that a supergravity description should exist at low-energy. It is difficult to make definite statements,  but some qualitative remarks
go as follows.

From a 4d perspective, assuming that the 4d SUGRA description is valid, one can use the general formulae of soft terms as in the beginning of the previous section. Alternatively, we can use higher-dimensional brane-antibrane
forces as a starting point of computing soft terms. At tree-level, we found qualitative agreement for distant (from the throat) D3 branes, which seem to fulfill an approximate sequestering,  broken 
by threshold effects that we commented around (\ref{m05}).   
Additional contributions come from small mass terms related to the contribution to supersymmetry breaking of the conifold field $Z$, which has an effect similar to 
an effective $(1,2)$ flux. As mentioned previously, it is also plausible that the warping is suppressing further the real values of these masses.  
  At one-loop level, 4d anomaly-mediated contributions to soft terms are present, consistent with a diffuse transmission of supersymmetry breaking across the bulk. Their higher-dimensional origin
 is not clear and it would be
  very interesting to  check them in detail from the higher-dimensional viewpoint, in particular the one-loop contribution to  the gaugino masses.  We hope to be able to return to this issue
  elsewhere. 

\section{Conclusions}
\label{Conclusions}

We continued the analysis of the KKLT moduli stabilization with antibrane uplift in the context of the Klebanov-Strassler warped compactification, taking into account the light complex structure
 field $S$ ($Z$) identified in \cite{Bena:2018fqc, Blumenhagen:2019qcg}.  
 Assuming the validity of the 4d supergravity description and using a manifestly supersymmetric formulation of the uplift via a nilpotent field on the antibrane, we worked out the vacuum structure and physical spectrum, confirming
 the potential destabilization of the KS throat. The minimal value of the needed flux (\ref{intro1}), combined with needed redshift for various physical purposes, translates into (relatively) large values
 of the localized D3 charge $Q_3^{\rm loc}$, beyond the usual perturbative values  $Q_3^{\rm loc} \leq 16$. Notice that this is not a surprise in itself. Indeed, the gravitational KS solution is valid for
 $g_s M \gg1 $, which is generically stronger than our destabilization limit $\sqrt{g_s} M \geq 6.8$.  However, using just the standard KS validity bound leads to weaker limits on  $Q_3^{\rm loc}$ by choosing
 maximal values ($g_s \sim \mathcal{O}(1)$) of the string coupling, which are reasonable at least in F-theory. The destabilization bound, on the other hand, leads to limits on  $Q_3^{\rm loc}$ which are
 independent on the string coupling and feature generically large numbers.  Our viewpoint in this paper is that, whereas it could be difficult to obtain such large localized D3 charges in perturbative type II strings,
  it is presumably possible in F-theory. 
In such F-theory models, however, the stabilization of a large number of complex structure moduli is challenging \cite{blaback}. 
   
 We studied the consequences of including the light complex structure modulus $Z$ in the low-energy description for the vacuum structure and phenomenology.
The effect of the antibrane uplift is that the vev of $Z$ is shifted such that its contribution to supersymmetry breaking is large, comparable with that of the nilpotent field. Secondly, whereas the 
conifold field is generically heavier than the volume modulus, justifying an integrating out procedure, this is not always the case and there is a condition that the flux and the other parameters  have to satisfy.  
Most of supersymmetry breaking is localized at the tip of the throat, but  a small amount of supersymmetry breaking is transmitted across the bulk far from the throat. 
We studied the effects of the supersymmetry breaking on observable sector fields, in particular on distant D3 branes. 
Neglecting threshold effects, we found 
 an approximate sequestering, with the only non-zero masses in the observable sector
are of $\mu$, $B \mu$ type and very small. Their appearance  could be due to the fact that the complex structure modulus $Z$ contributes to supersymmetry breaking, which acts effectively as  an $(1,2)$ flux.   
Comparing the results from a four-dimensional supergravity perspective and from a higher-dimensional one, one finds qualitative agreement at tree-level. Threshold effects  (\ref{m05})  are expected to add 
other contributions breaking sequestering  \cite{Choi:2005ge,Berg:2010ha}. It would be interesting to understand the origin of the term $\Delta K$ in  (\ref{m05}) in the nilpotent formalism, needed in order to 
match non-holomorphic soft scalar masses with previous results.

It would be clearly interesting to investigate  further one-loop soft terms from four dimensional and higher-dimensional perspective and check their compatibility. 
It would also be interesting to study from a similar viewpoint other models of moduli stabilization and check if the potentially dangerous  throat destabilization still exists. The existence of a critical 
value of the flux to avoid destabilization could be addressed directly in the dual gauge theory of the KS  throat. Another direction to investigate
is the search of alternatives, to the $\overline{D3}$ antibrane,  uplifts of the vacuum energy. 
Finally, we commented on the dS swampland conjecture \cite{obied}. It would be interesting to investigate further this refined model of moduli stabilization from the viewpoint of the  other swampland conjectures 
\cite{swampland-brotherhood-graveyard}  (for
an extensive review and references, see \cite{eran}). 

\vskip 1cm
\noindent {\bf Note added:}  While this paper was completed, the paper \cite{Randall:2019ent} appeared, which has some overlap with ours and interpret our normalized field $Y$ of Section~\ref{sec:4D} with the radion
from a 5d perspective. We thank Lisa Randall for discussions and for sharing her preliminary draft.

\vskip 1cm
\noindent {\bf Acknowledgements:} 
We are grateful to Iosif Bena and Mariana Gra\~na for collaboration in an earlier stage of this work and many important discussions. We would like to thank Ralph Blumenhagen, Arthur Hebecker, Luca Martucci and Lisa Randall for very helpful discussions and correspondence. 
This work was supported in part by the ANR grant Black-dS-String ANR-16-CE31-0004-01 
and the ERC Consolidator Grant 772408 ``Stringlandscape''.
 

\hskip 1cm

\appendix
\noindent
\section{Appendix}

As well-known, the 4d supergravity action does  not depend separately on $W$ and $\cal K$, but on  the function $G = {\cal K} + \log |W|^2$. Some  
 useful formulae for a K\"ahler space, for which $G_{I \bar J} = K_{I \bar J} = \partial_I \partial_{\bar J} {\cal K}$,  used in the text and in particular to evaluate  (\ref{soft3})  are 
\begin{eqnarray}
&& G^I = G^{I \bar J} G_{\bar J} \ , \ \Gamma_{IJ}^K = G^{K \bar M} \partial_I G_{J \bar M} \quad  , \quad \nabla_I V_J = \partial_I V_J -    \Gamma_{IJ}^K V_K \ ,   \nonumber \\
&&   R_{I \bar J K \bar L} = \partial_K \partial_{\bar L} G_{I \bar J} - G^{M \bar N} \partial_K G_{I \bar N}
\partial_{\bar L}  G_{M \bar J}    \  .  \label{a1}
\end{eqnarray}

In the case of our effective action (\ref{m2}), the nontrivial components of the connections are ($i,j$ denote matter fields indices in what follows)
\begin{equation}
\Gamma_{i \rho}^j = - \frac{1}{r} \delta_i^j \quad , \quad \Gamma_{i y}^j =  \frac{\bar y}{r} \delta_i^j \quad , \quad \Gamma_{i j}^\rho = 4 z_{ij} {\bar y}^3  \quad , \quad \Gamma_{i j}^y = 6 z_{ij} {\bar y}^2
  \  .  \label{a2}
\end{equation}
By using the metric components and  (\ref{a1}),  (\ref{a2}), one can easily verify
\begin{equation}
R_{i {\jbar} \alpha \bar \beta}  =  \frac{1}{3}  G_{i \jbar}  G_{\alpha \bar \beta}    \  ,  \label{a3}
\end{equation}
where $\alpha, \beta = X,Y,\rho$ and $i,j$ are matter fields indices. 
In the presence of the correction to the K\"ahler potential (\ref{m05}) breaking sequestering, there is a new correction to the Riemann tensor, which becomes, at first order in the correction, 
\begin{equation}
R_{i {\jbar} \alpha \bar \beta}  =  \frac{1}{3}  \left( 1 + \frac{D}{r^2} \delta_{\alpha X}  \delta_{\bar \beta \bar X}  \right) G_{i \jbar}  G_{\alpha \bar \beta}  \  .  \label{a4}
\end{equation}
Analogously, for the correction in (\ref{m6}) , the Riemann tensor  becomes
\begin{equation}
R_{i {\jbar} \alpha \bar \beta}  =  \frac{1}{3}  \left( G_{\alpha \bar \beta} - \frac{2 \xi' |Y_0|^{4}}{N d^6} \delta_{\alpha X}  \delta_{\bar \beta \bar X}  \right) G_{i \jbar}  \  .  \label{a4}
\end{equation}


\providecommand{\href}[2]{#2}\begingroup\raggedright\endgroup

\end{document}